# Fracture imaging within a granitic rock aquifer using multiple-offset single-hole and cross-hole GPR reflection data

Caroline Dorn[1], Niklas Linde[1], Joseph Doetsch[2], Tanguy Le Borgne[3], Olivier Bour[3]

[1]Institute of Geophysics, University of Lausanne, Lausanne, Switzerland;

[2]Institute of Geophysics, ETH Zurich, Zurich, Switzerland;

[3]UMR Geosciences Rennes 6118, University of Rennes, Rennes, France.

ABSTRACT

The sparsely spaced highly permeable fractures of the granitic rock aquifer at Stang-er-Brune (Brittany, France) form a well-connected fracture network of high permeability but unknown geometry. Previous work based on optical and acoustic logging together with single-hole and cross-hole flowmeter data acquired in 3 neighboring boreholes (70-100 m deep) have identified the most important permeable fractures crossing the boreholes and their hydraulic connections. To constrain possible flow paths by estimating the geometries of known and previously unknown fractures, we have acquired, processed and interpreted multifold, single- and cross-hole GPR data using 100 and 250 MHz antennas. The GPR data processing scheme consisting of time-zero corrections, scaling, bandpass filtering and F-X deconvolution, eigenvector filtering, muting, pre-stack Kirchhoff depth migration and stacking was used to differentiate fluid-filled fracture reflections from source-generated noise. The final stacked and pre-stack depth-migrated GPR sections provide high-resolution images of individual fractures (dipping 30-90°) in the surroundings (2-20 m for the 100 MHz antennas; 2-12 m for the 250 MHz antennas) of each borehole in a 2D plane projection that are of superior quality to those obtained from single-offset sections. Most fractures previously identified from hydraulic testing can be correlated to reflections in the single-hole data. Several previously unknown major near vertical fractures have also been identified away from the boreholes.

1 Introduction

The hydraulic response of fractured rock aquifers is largely governed by the spatial organisation of permeable fractures. Identifying and characterizing individual permeable fractures or flow paths at the local field-scale (1-100 m) is an important and largely unresolved research goal for the hydrological and geophysical research communities (Long et al., 1996; Day-Lewis et al., 2006). Fractured rock masses are used worldwide, among others, for water supply purposes (e.g., Caruthers and Smith, 1992), as host rocks for environmentally hazardous waste (e.g., Mair and Green, 1981) and their characterisation is necessary in rock fall prone areas (e.g., Spillmann et al., 2007). Single-hole ground penetrating radar (GPR) is a powerful technique to map potential permeable fractures and fracture zones away from boreholes and at large depths in relatively resistive rock (e.g., Olsson et al., 1992; Hollender et al., 1999), whereas surface GPR can be very useful down to some 10-20 m depth in sparsely fractured crystalline and metamorphic rock (Grasmueck, 1996). Borehole or surface-based time-lapse GPR experiments carried out during saline tracer tests, or combined interpretations of hydraulic data and borehole or surface-based GPR may identify larger m-scale fractures that are permeable and significantly contribute to the local fluid flow (Day-Lewis et al., 2003, 2006; Talley et al., 2005; Tsoflias et al. 2001; Tsoflias and Becker, 2008).

Most previous work using single-hole GPR reflection data have used only one single-offset data from lower frequency antennas, such as 60 MHz, without migrating the data (see Spillmann et al. (2007) for an example of migrated single-offset high-frequency data). Hollender et al. (1999) illustrated that multiple-offset data can significantly improve the resolution of single-hole GPR sections. We expect that using high-frequency (100 and 250 MHz) multiple-offset data together with advanced processing will allow us to further improve the results of single- and cross-hole GPR investigations compared with those present in the literature. We also want to investigate to what extent cross-hole radar reflections complement single-hole reflection data. More importantly, the field-based results are expected to provide critical information about fractures that cannot be obtained from hydrological investigations alone.

Our field site is a well-studied hydrological research site located in a crystalline aquifer in Brittany (Stang-er-Brune, Fig. 1a) in which Le Borgne et al. (2007) performed extensive hydrological testing and borehole logging. They concluded that the local conductive fracture network is dominated by only a few well-connected fractures (i.e. only 3-5 such fractures intersect a borehole over its entire length of ~90 m). High-resolution borehole images of the transmissive fractures show that these fractures have predominantly dip angles between 30° and 70°. The geometry of the hydrological connections between fractures are unknown. This is illustrated by apparent connections dipping up to 80° and that none of the permeable fractures appear to cross more than one borehole. The geometry of the permeable fracture network remains largely unknown as borehole data only provide detailed information in the close vicinity of the boreholes, whereas the single-hole and cross-hole flowmeter data provide information about connections, but not their geometry (Le Borgne et al., 2006).

To better understand the geometry of potential flow paths by imaging single fractures at the site, borehole GPR experiments were carried out in June 2009 to image single fractures up to some 20 m from the boreholes. We acquired multifold single-hole and cross-hole GPR data in three boreholes (B1-B3, see Fig. 1) using 100 MHz and 250 MHz antennas. This data are here used to determine the size, dip angles and to constrain the possible orientations of single fractures, especially for those that have not been previously identified, as they do not intersect the boreholes. We expect the strongest recorded GPR reflections to originate from the major open water-filled fractures (Tsoflias et al, 2008) and at the ~40 m deep contact between mica schist and granite (Fig. 1).

In this contribution, we describe the processing of data acquired in borehole B1 (single-hole) and within the borehole-plane B1-B2 (cross-hole). We then present the final migrated sections including the boreholes B1-B3. The results are then interpreted together with available hydrological and borehole logging data.

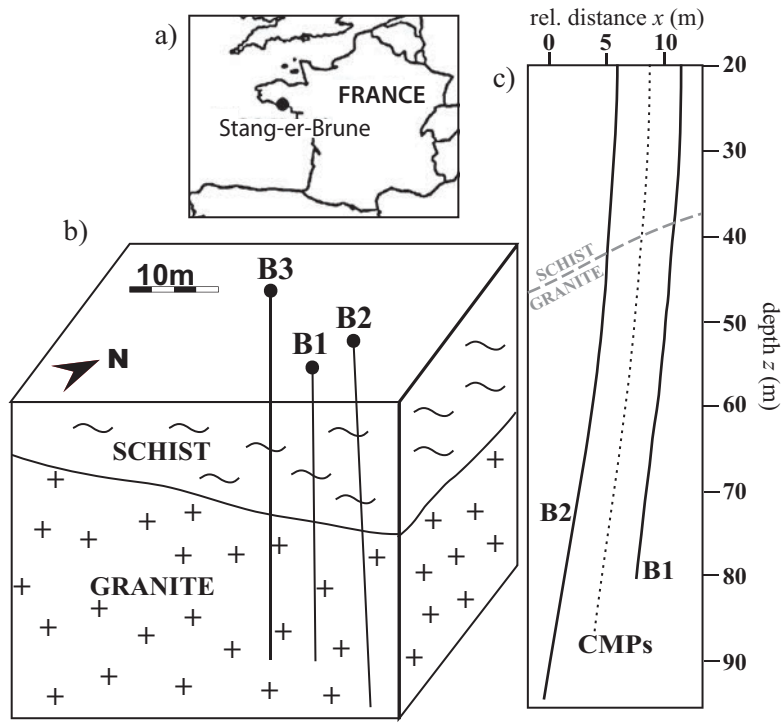

*Fig. 1: (a) Location of Stang-er-Brune, France. (b) Geological model of the field site showing the 30° dipping contact between mica schist and underlying granite. (c) 2D projection of the borehole geometry of B1, B2 and cross-hole CMP (common midpoint) locations. At z >50 m there is a relative dip between B1 and B2 of ~3°.*

## 2 General setting of the crystalline aquifer

Our field-site in Brittany (Stang-er-Brune), France, is located 3 km west of the main groundwater-pumping site of Ploemeur and is part of a long-term hydrological research observatory (http://hplus.ore.fr). The 3 deeper boreholes installed at the site reach depths from 80 to 100 m (water table during acquisition at ~1.5 m depth). The borehole deviations are up to 6° from the vertical as estimated by a deviation probe using a three-axis fluxgate magnetometer for bearing and a three-axis accelerometer for inclination. This tool has an estimated uncertainty of 0.5°, which leads to an expected standard deviation of 0.9 m in the horizontal coordinates at 100 m depth. The characterization of fractures and the geology of the site are known from borehole coring (B1) and optical, acoustic, gamma-ray and electrical geophysical logs that are discussed in detail in Le Borgne et al. (2007) and Belghoul (2007). At a regional scale, the geology is characterised by low-porosity granite overlain by micaschists, with a contact zone dipping ~30° towards north (Touchard, 1999; Le Borgne et al., 2006, Ruelleu et al., 2010). The contact zone is relatively complex and consists of alternating deformed granitic sheets and enclaves of micaschists, pegmatite and aplite dykes (Ruelleu et al, 2010). At the Stang-er-Brune site, we observe a part of this contact zone with micaschists in the first 30 to 40 meters overlying the Ploemeur granite. The formation is highly transmissive with overall transmissivity over the depth of each borehole varying around $10^{-3}$ $m^2/s$. This high transmissivity implies a strong connectivity of the permeable fractures that is probably related to the contact zone between the intrusive granite and the overlying micaschists.

**3 Multifold data acquisition**

Single-hole GPR data were acquired in all three boreholes (B1-B3) in common offset sections using 16 (4) different transmitter-receiver separations for the employed 100 MHz (250 MHz) antennas with a depth sampling of 0.15 m (0.1 m). The antenna offsets were equally distributed in the range of 2.3-11.3 m (100 MHz antennas) and 1.8-7.8 m (250 MHz antennas). The dominant frequencies are around 70 (140) MHz for the data acquired with the 100 (250) MHz antennas.

The cross-hole GPR data were acquired for all three borehole planes (B1-B2, B1-B3, B2-B3) in common transmitter gathers. The cross-hole planes of B1-B2 were acquired with 250 MHz antennas, whereas 100 MHz antennas were used for B2-B3 and B1-B3. Transmitters were spaced every 0.5 m and receivers every 0.1 m resulting in nominal 30-fold data at a common-midpoint (CMP) spacing of 0.125 m. The acquired offsets were restricted to those for which reasonable signal-to-noise ratio data could still be obtained.

To protect and to center the antennas in the boreholes, we attached two self-made plastic packers to each antenna, with a slightly smaller diameter than those of the boreholes (10.5 cm). Antennas were first positioned by transforming distances along the borehole measured with a trigger wheel into depths with the help of the deviation logs. Differential GPS was used to measure the top of the borehole casings. All depths are given relative to the top of the B1 casing. Time-zero measurements were performed when changing the antenna separation (single-hole) or after measuring 40 source-gathers (cross-hole).

## 4 Data processing

Significant ringing in the 250 MHz raw data and the dominance of the direct wave at early times, together with positioning uncertainties and radar wavespeed variations, resulted in numerous processing-related challenges. The processing scheme (Table 1) addressed these issues and resulted in high-quality GPR images of the surrounding fractured rock matrix. Even if the individual processes are standard in seismic imaging (Yilmaz, 2001), it was necessary to adapt the processing and the parameters to address the specific characteristics of the borehole GPR data.

We begin the description by first explaining the progressively improved radar wavespeed and borehole trajectory estimates before we outline the specific processing schemes used for single-hole and cross-hole data.

In the following, dips of reflectors are always given with respect to the surface assuming vertical boreholes if not mentioned differently. Note that shallow dips (relative to the surface) refer to steep dips of features relative to the subvertical borehole trajectories.

*Table 1: Main processing steps of single- and cross-hole GPR data.*

| Representative Figure | **Single-hole data processing** | **Cross-hole data processing** | Representative Figure |
|---|---|---|---|
| | Static corrections | Static corrections | |
| 2a → | AGC | AGC | |
| | Bandpass filter | Bandpass filter | |
| | F-X deconvolution | F-X deconvolution | |
| | Eigenvector filter | Eigenvector filter | |
| 2b → | First-break mute | First-break mute | |
| 2c, 3a → | Prestack depth migration | Dip-decomposition dip-moveout (includes stack) | ← 7b |
| | Custom mute | | |
| 3d → | CMP Stack | Kirchhoff depth migration | ← 7c |

### 4.1 Radar wavespeed estimation and borehole trajectory estimation

The single- and cross-hole data were used together to define average radar wavespeed functions *v(z)* and refine the estimated borehole trajectories. Isotropy of the radar wavespeed was assumed on the wavelength scale to allow for comparison of the single- and cross-hole data. This assumption is reasonable as granite is an igneous rock of low porosity having negligible intrinsic anisotropy.

An initial radar wavespeed model was obtained from the single-hole data by fitting a straight line to the slope of the corrected direct wave (which travels at the wavespeed of the rock in the vicinity of the borehole) travel times vs. transmitter–receiver distances for each CMP location. This procedure allows us to avoid static errors, resulting for example from the assumption that a finite-length antenna can be represented as a point source or by neglecting the delay of the direct wave in the water filled space between the antenna and the borehole. At $z > 40$ m, the differences between the radar wavespeed functions from all three boreholes are 2.5 % on average.

The first radar wavespeed estimates were subsequently refined by traveltime tomography. The mean of the 1-D radar wavespeed functions of each borehole pair for which tomography was carried out was used as initial and reference models in the tomography (green line in Fig. 2a). The tomography was mainly carried out to improve the borehole positioning and to evaluate if the radar wavespeed estimates from the cross-hole data correspond with those determined in the vertical direction (i.e., assumption of isotropy). For the tomography, we followed the inversion procedure of Linde and Doetsch (2010) using a stochastic regularisation operator based on an exponential isotropic covariance model with integral scales of 0.5 m aiming at fitting the data to an error level of 1.0 ns. The integral scale corresponds to the distance at which the spatial correlation has decreased to $1/e$. To decrease the sensitivity to noise in the data and to image sharp radar wavespeed variations we used a formulation based on iteratively reweighted least-squares (Farquharson, 2008).

Initial inversions of the cross-hole traveltimes in the plane B1-B2 resulted in suspiciously high radar wavespeeds at $z > 50$ m (blue line in Fig. 2a) compared with the initial mean radar wavespeed function B1/B2 (green line in Fig. 2a) indicating that the borehole spacing is smaller than indicated by the deviation logs. This inversion reached a final RMS error of 1.84 ns after 10 iterations. To improve the results it was necessary to correct the antenna positions in both the vertical and horizontal directions. For each transmitter gather, the receiver array was shifted vertically to minimise offset-correlated behaviour in the residuals. This correction (max. 0.5 m) was in all cases increasing with depth and helped to reduce small-scale variations in the resulting radar wavespeed models. Twisting of the antenna cables might have caused these errors.

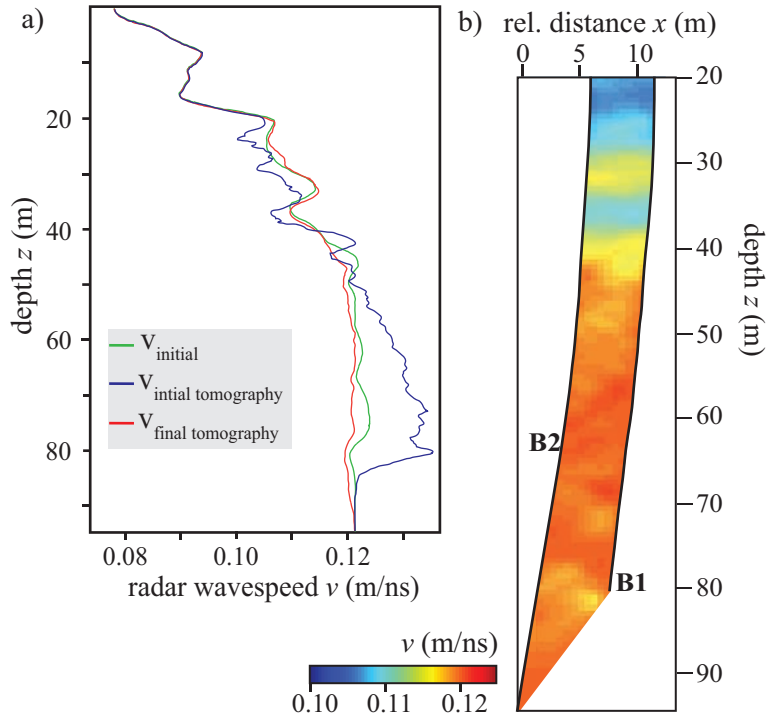

*Fig. 2: (a) 1-D radar wavespeed estimates for the plane B1-B2: (blue) after tomographic inversion of the cross-hole traveltimes using the initial borehole geometry and the initial radar wavespeed function (green) as starting and initial model, (red) after using the same tomographic inversion scheme but with final geometrical corrections applied. (b) Tomogram derived from first arrivals using the initial radar wavespeed function (a, green line) as starting and reference model and with final geometrical corrections.*

To further constrain the distances between B1 and B2, we picked the observed prominent reflections originating from the adjacent borehole in the depth-migrated single-hole data (using the initial radar wavespeed function shown as green line in Fig. 2a). We corrected for the relative borehole geometry given the shape of the picked reflector distances. To correct for significant borehole deviation errors for the other cases where we could not identify reflections from the adjacent boreholes (B2-B3 and B1-B3), we tested horizontal correction factors that linearly varied with depth until the residuals (after the first iteration) have the lowest correlation possible.

The geometrical corrections and the subsequent inversion of the B1-B2 data resulted in the tomogram shown in Fig. 2b with an rms error of 1.05 ns after 10 iterations. As expected, the traveltime tomography provides no information about individual fractures but images large-scale radar wavespeed variations and trends. The final radar wavespeed model plotted in red in Fig. 2a is based on the horizontal average of the tomographic model in regions with dense ray-coverage ($z$ = 20 to 80 m) and the initial model outside this region. The final radar wavespeed estimates are mostly in the range of 0.09±0.002 m/ns (mica schist) to 0.12±0.002 m/ns (granite). Given the small lateral variations in the tomogram (Fig. 2b), it appears that the assumption of a 1-D radar wavespeed function holds well if all applied corrections prior to the inversion are correct. These 1-D radar wavespeed functions were used to migrate the single- and cross-hole data.

**4.2 Single-hole processing**

**4.2.1 Processing challenges**

The GPR single-offset section (B1, 250 MHz antenna) in Fig. 3a illustrates some of the data characteristics. Source-generated noise (N) or rather poor coupling within the antenna-borehole-rock system creates ringing effects parallel to the direct wave (D) but at later times. The ringing is most critical for the 250 MHz data, where it is superimposed on early reflections (R). The reflectivity pattern varies along the borehole and there is an abrupt increase in signal-to-noise ratios at approximately 40 m depth below the schist.

A general problem in single-hole GPR imaging with standard commercial omni-directional antennas (directional borehole antennas exists (e.g., Slob et al, 2010) but are not widely used) is that the orientation of the fractures cannot be determined using data from one borehole alone. The data carries no information about the direction at which a reflection wavefront arrives at the borehole. Our processing and migration is therefore carried out under the assumption that the radar wavespeed varies only in the vertical direction (1-D radar wavespeed function $v(z)$, where $z$ is depth). We showed in section 4.1 that the 1-D radar wavespeed assumption appears to hold well within the granite, which is of primary interest in this study as it is the host rock of the most permeable fractures.

Borehole logging data (optic and acoustic) and analysis of the retrieved core of B1 indicate that the shallowest dips of transmissive fractures are in the range of 30° (except one fracture dipping 15°; Le Borgne et al., 2007). Using migration methods to image such subhorizontal dips (steep dips with respect to the observation line) require an accurate radar wavespeed model.

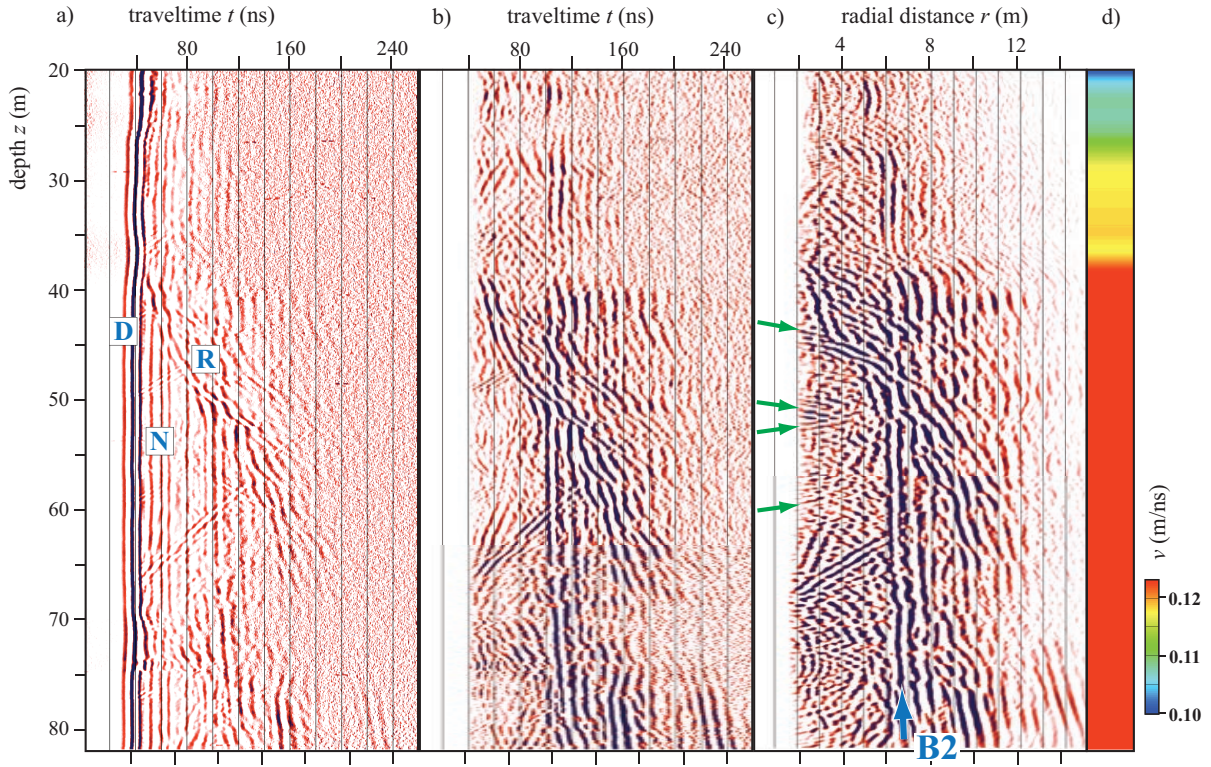

***Fig. 3:*** *Results of pre-stack processing applied to a typical common-offset section of B1 (3.8 m offset, 250 MHz antennas). (a) Section with applied static corrections and an AGC of 70 ns window length. Letters indicate regions dominated by: D – direct wave; N – source-generated noise (ringing); R - reflections. (b) As in (a) but after application of bandpass filter, F-X deconvolution, eigenvector filter and custom mute (see text). (c) As in (b) but after pre-stack depth migration (the axis aspect ratio r:z is 2:1). The blue arrow in (c) indicates reflections generated from the adjacent borehole B2; the green arrows refer to features discussed in the text. (d) 1-D wavespeed model for depth migration in (c).*

### 4.2.2 Pre-stack time-domain processing

Table 1 lists the main single-trace and multi-trace filter and deconvolution tools that were applied to the single-offset data. All parameters within the processing sequence were chosen to account for the different frequency contents of the 100 and 250 MHz data.

Corrections were carried out to account for inaccuracies in the data acquisition sampling frequency (c.f. Hollender et al., 1999), drifts in the time-zero and geometrical positioning errors. After appropriate scaling (Fig. 3a), reflections (R) had to be enhanced and separated from the direct wave (D) and ringing noise (N), resulting in a common single-offset section as shown in Fig. 3b. Bandpass filtering (30-270 MHz for the 250 MHz data, 20-130 MHz for the 100 MHz data) removed low- and high frequency noise from the data. F-X deconvolution increased reflection coherency by reducing random noise without introducing noticeable artefacts. For the 250 MHz data, an eigenvector filter was applied in a window around the direct wave to remove energy parallel to the direct wave mainly containing ringing noise (N). Residual energy that appears earlier than the direct wave is muted afterwards.

**4.2.3 Pre-stack depth migration and stacking**

Pre-stack Kirchhoff depth migration of single-offset sections provided high-quality image reflections with overlapping features and a wide dip angle range (Fig. 3c). The migration radar wavespeeds used are in the range of 0.09 m/ns (mica schist) to 0.12 m/ns (granite). In section 4.2, we describe how we estimated the radar wavespeed function.

We used a migration method that computes first arrivals through an implicit Eikonal solver. The amplitudes are neither calculated nor meaningful as the processing is based on a monopole (point) radiator, which does not correspond well with the dipole-like radiation characteristics of borehole antennas. In regions without significant radar wavespeed variations (40-100 m), subhorizontal dipping features (down to 30°) are well preserved after migration (green arrows in Fig. 3c). Migration artefacts are likely to be present around $z = 40$ m (Fig. 3c) since the 1-D radar wavespeed model does not account for the 30° dip angle of the mica schist-granite contact. Apart from this region, extensive comparisons of migrated sections using a variety of radar wavespeed models indicate that the sections are free of major migration artefacts.

Stacking the migrated single-offset sections of the 100 MHz data significantly improved the quality compared to individual single-offset sections (compare Figs. 4a and c). Close to the borehole, subhorizontal dipping features (e.g., ~30°) are less well imaged at larger antenna offsets (see arrows in Figs. 4b and c). Further away from the borehole, reflections from subhorizontal dipping features of limited extent are best recognised at larger antenna offsets (see ellipses in Figs. 4b and c). The image quality increases for larger radial distance $r >$ ~8 m by adding information from larger offset sections, but decreases for $r <$ ~6 m. To avoid sub-optimal resolution close to the boreholes in the final image, we applied an offset-dependent top mute prior to stacking (Fig. 4d); the larger the offset the longer the applied top mute.

For the 250 MHz data we stacked the three largest offset sections, since the shortest offset (1.8 m) sections were highly contaminated by ringing effects and did not contribute significantly to the stacked images. The major improvements of stacking the 250 MHz data came from the additional offset information and less from an increased signal-to-noise ratio, given that only a few offsets were used.

We do not image any features at distances $r < 2$ m because of the dominance of the direct wave at early times and its subsequent removal, which also tends to remove superimposed reflections at early times. This complicates direct comparisons with the borehole logging data.

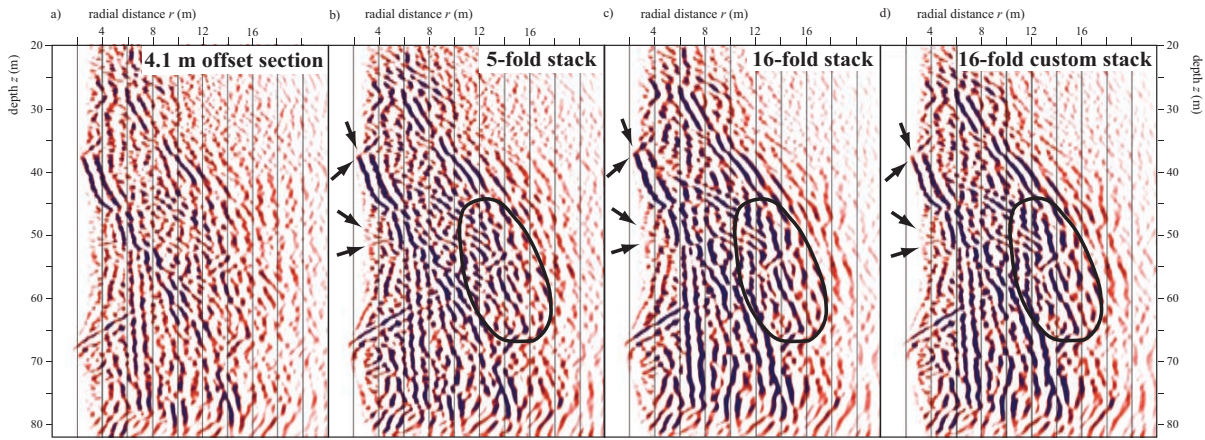

*Fig. 4:* Illustration of the improved signal-to-noise ratios offered by stacking the migrated images of B1 (100 MHz data). (a) Single-offset section with 4.1 m antenna separation. (b) Stack of the 5 shortest antenna separations, (c) all 16 offset sections, and (d) as in (c) but with a custom mute applied for each offset section before stacking. Note that subhorizontally dipping events at small radial distances are best imaged in (b) and (d) (see arrows), whereas at larger radial distance they are best imaged in (c) and (d) (see ellipses).

### 4.3 Cross-hole processing
### 4.3.1 Challenges

Common cross-hole processing tools such as single-trace mapping techniques (Khalil et al., 1993; Lazaratos et al., 1995) that focus on layered structures can only handle horizontal to sub-horizontal reflectors correctly. The complex spatial distribution of fractures at the site and the cross-hole acquisition geometry necessitate an approach that can accurately consider all reflector dips.

Unlike the cylindrical symmetry of the single-hole data, reflections in the cross-hole sections can originate from any point on an ellipsoid around a respective CMP location (Fig. 5). Some analysis is therefore needed to better understand how to migrate such data and interpret the results in terms of possible dip angle and distance ranges. For this geometry, any signal travelling between the two boreholes B1 and B2, being reflected on a 90° dipping plane at $y = P_y$ (Fig. 5a) cannot be distinguished from any other reflection on the ellipse described by the two main axes $P_x$ and $P_y$. In the depth-migrated image this plane would show up at a shorter distance $x = P_x$. A reflector plane at $y = P_y$ dipping 60° away from the borehole (see solid red lines in Figs. 5a-c) is imaged in the depth- migrated section as a curved feature (see corresponding dashed red lines in Figs. 5a-c).

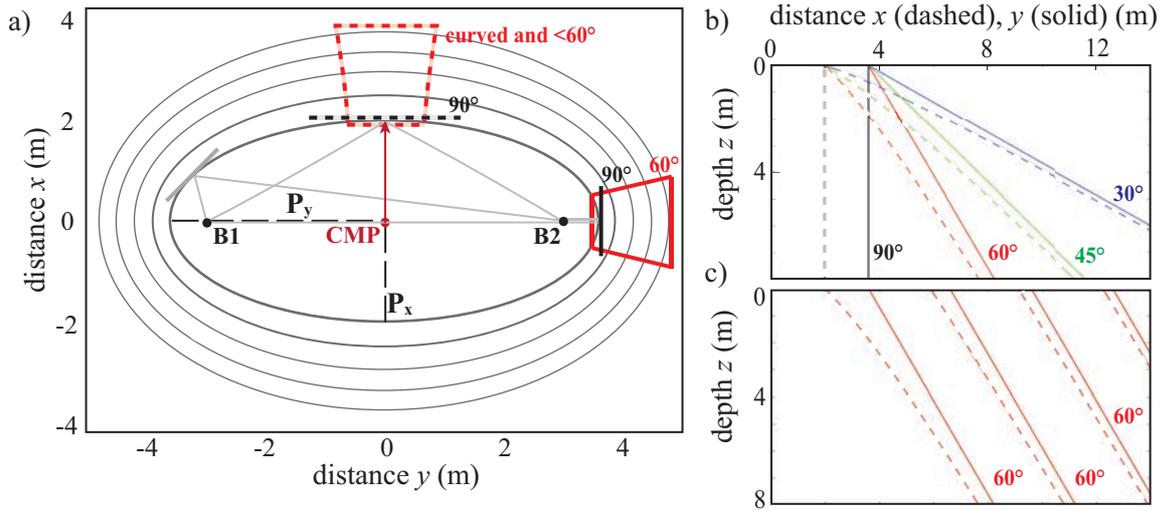

*Fig. 5: (a) 2D-schematic of elliptic distance projection in cross-hole reflection imaging showing boreholes B1 and B2 from a bird's eye view. A 90° dipping reflector at a distance x = $P_x$ in the depth-migrated GPR image can originate from anywhere on an ellipse described by the two main axes $P_x$ and $P_y$ surrounding the CMP. (b) and (c) Dip and distance representations in the depth migrated images. The solid lines describe tangential reflector planes striking perpendicular to the inter-borehole plane and dipping at angles 30°, 45°, 60° and 90° (see solid lines in (a) of both 90° and 60° dipping planes). The dashed lines describe how they would appear in a depth-migrated image (see dashed lines in (a) of plane representations).*

To simulate the geometrical effects of the examples described above (90° and 60° dipping reflector), we have computed synthetic seismograms for the B1/B2 borehole geometry using Bohlen's (2002) 3D viscoelastic finite difference modelling code (Fig. 6). An alternative would have been to use a 3D GPR code (e.g., Giannopoulos, 2005). The use of synthetic seismograms can help to explain the GPR reflection geometry, but not the GPR amplitudes. The modelled and depth-migrated data confirms our geometrical assumption of an elliptic projection. Calculated and observed distances in the depth-migrated image fit well.

Fig. 6b confirm that planar features might appear curved in the migrated sections.

Due to the removal of the direct wave that strongly contaminated early reflections, the earliest reflections that are possible to image originate from an ellipse (see ellipse in Fig. 5a) in a range from 2 to 4 m away from the CMP.

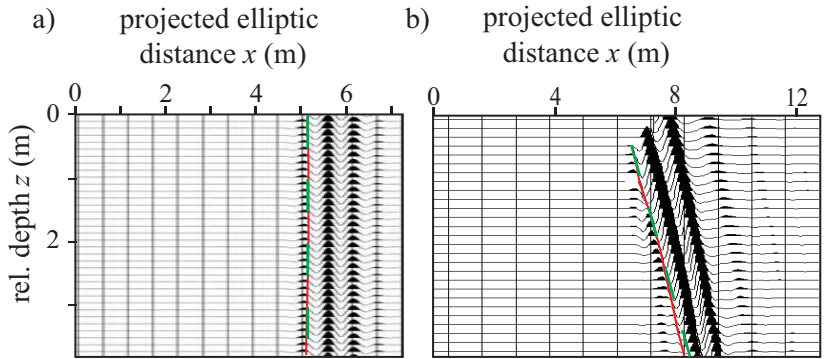

*Fig. 6: (a) Modelled depth-migrated seismic data for a borehole geometry as in Fig. 5a and a 90° dipping reflector plane at y = 5.9 m away from the CMP. The picked (red line) and predicted distances (green line) are both x = 5.1 m. (b) As for (a), but for a modelled reflector plane dipping 60° away from the surface.*

### 4.3.2 Processing

The pre-processing of the cross-hole data was similar to that of the single-hole data (section 4.1) but was followed by dip decomposition DMO and depth migration (Table 1). The NMO stack (Fig. 7a) performed with the dip-independent stacking wavespeeds (Fig. 2a, red line), reveals some structural complexity but fails to clearly image subhorizontal dipping reflectors superimposed on subvertical features. We treated the preprocessed data with a dip-moveout (DMO) algorithm based on Jakubowicz (1990) that distinguishes and processes events on the basis of dip, with angles between 0° and 90° discretised into 45 different values. For each of the 45 dip angle values, the CMP gathers were NMO corrected using appropriate velocities estimated from the dip-independent wavespeeds based on standard formulas and then stacked. Finally, all 45 dip-filtered stacks were summed together to form a DMO-corrected stack (Fig. 7b). Compared to the NMO stack, the DMO process improved signal-to-noise ratios throughout the section and conflicting dips are better imaged (see rectangles in Figs. 7a and b). Other tested DMO algorithms did not sufficiently image subhorizontal dips (e.g., common-offset F-K DMO).

The migrated section (Fig. 7c) was obtained by a post-stack Kirchhoff depth migration for steep dips (in this context *steep* refers to dips with respect to the observation line, Table 1) using the same implicit eikonal solver as for the single-hole data. This allowed us to obtain images that were free of major artefacts. We decided to use partial pre-stack migration followed by post-stack depth migration because it showed the least migration artefacts in comparison with pre-stack-migration schemes performed on the modelled cross-hole seismograms (see also Fig. 6).

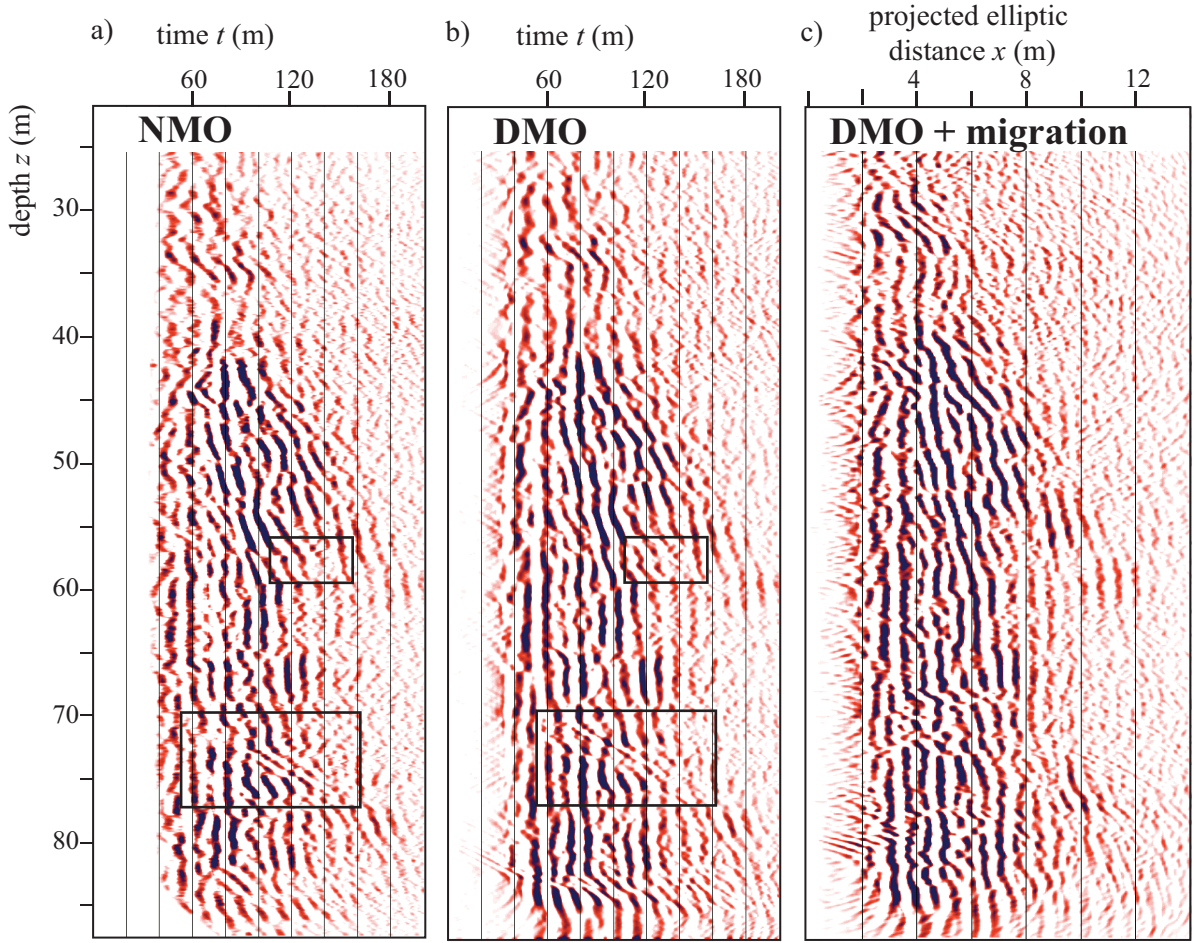

*Fig. 7: Post-stack processed cross-hole section for B1-B2. (a) NMO stack, (b) dip-decomposition DMO stack and (c) post-stack Kirchhoff depth-migrated version of (b) based on radar wavespeed model shown in Fig. 2a (red line). Black rectangles refer to features discussed in the text.*

## 5 Results

### 5.1 Single-hole GPR images

The final stacked and migrated sections of B1-B3 show several linear reflections (dipping 30-90°) located at 2-14 m and 2-20 m radial distance for the 250 MHz (Fig. 8) and 100 MHz (Fig. 9) data. These reflections are expected to mainly originate from individual fractures and fracture zones. The sections obtained from the 100 and 250 MHz antennas are comparable as most prominent features in the 250 MHz section are also represented in the 100 MHz section. The higher resolution of the 250 MHz data at $r < 8$ m allows more structural details to be imaged, especially for subhorizontal dipping features. The change from low to high reflectivity at $z = 38$ m in B1 ($z = 42$ m in B2, $z = 35$ m in B3) is related to the higher attenuation in the more conductive overlying mica schist compared to the granite.

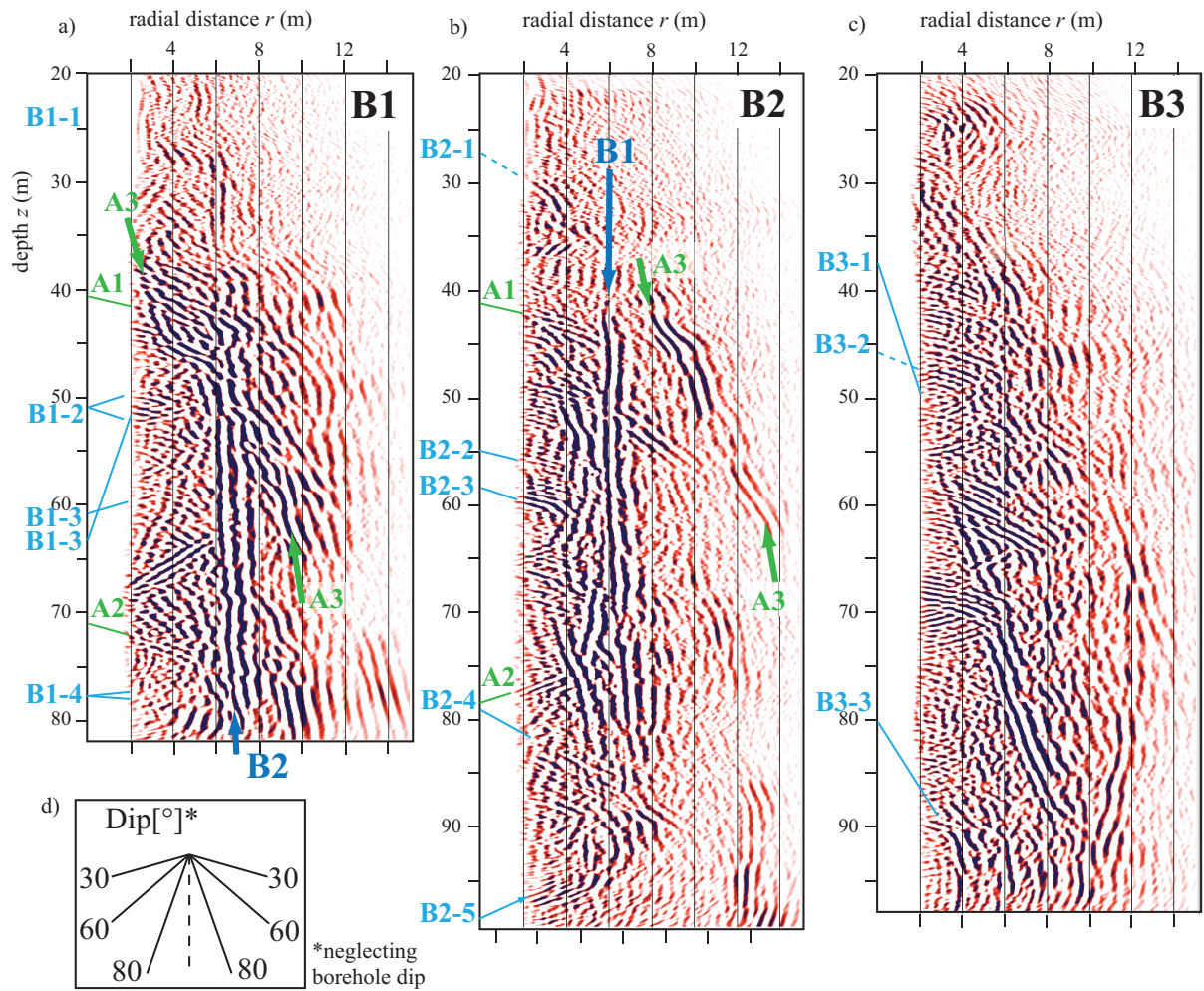

*Fig. 8: Processed and pre-stack depth-migrated single-hole GPR data acquired with the 250 MHz antenna in (a) B1, (b) B2, and (c) B3. Blue letters refer to transmissive fractures observed in the boreholes from optical logs and flowmeter tests (Le Borgne et al., 2007), green letters refer to additional features discussed in the text. The radial distance r indicates distances away from the respective boreholes. Note the lack of information at r < 2 m due to the direct wave removal. (d) Dip angles corresponding to the axis aspect ratio r:z of 2:1. Note that a given dip (0-90°) can be imaged with two different dip directions (e.g., A2).*

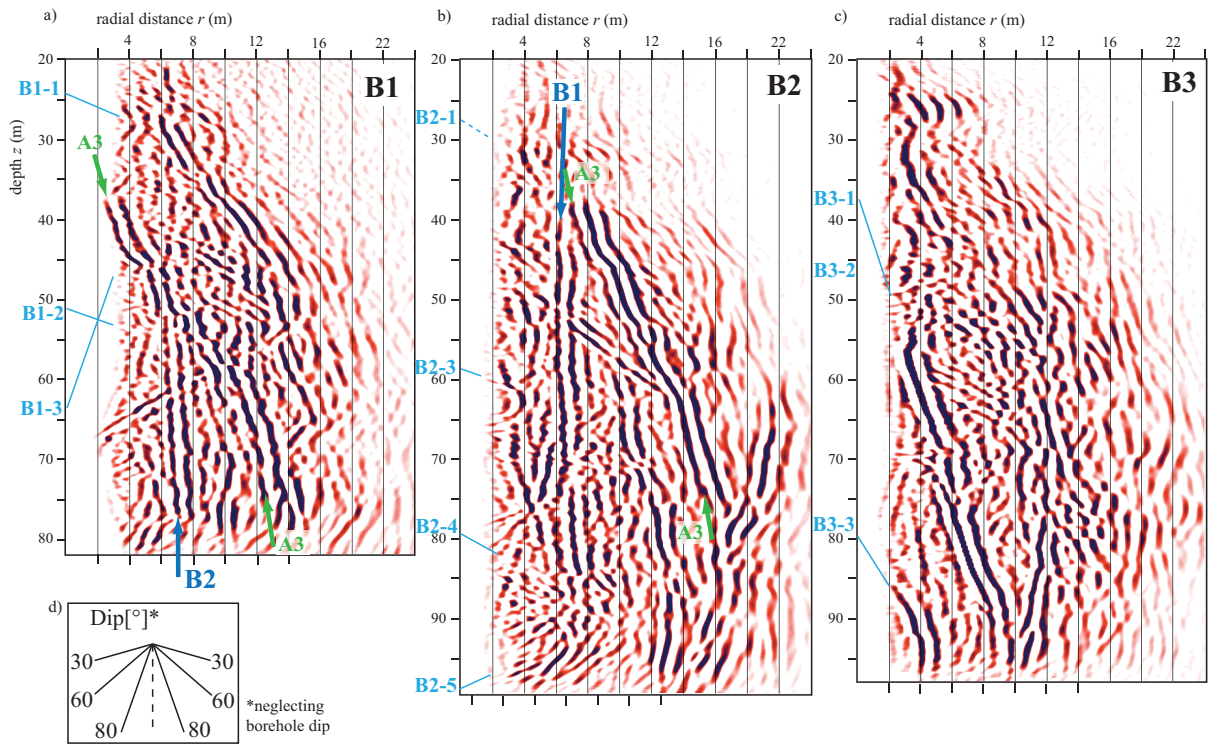

***Fig. 9:*** *The same borehole sections as in Fig. 8 but for the processed and pre-stack depth-migrated GPR data that were acquired with the 100 MHz antenna. Section (a) is identical to Fig. 4d.*

The prominent sub-vertical reflections at $r \approx 6$ m (arrows B2 and B1 in Figures 8 and 9) correspond to the neighbouring borehole. Identical reflector planes can be recognised in both B1 and B2 GPR sections as having similar dips and corresponding depth ranges. We can also distinguish reflections originating from outside the borehole plane B1-B2 from reflections generated between the boreholes. Reflections originate from outside the plane B1-B2 if the imaged reflections have similar dip angle and dip direction for both single-hole sections (see A1 in Fig. 8). Reflections originate from within the borehole plane if they have similar dips but the dip direction appears different comparing both single-hole sections (see A2 in Fig. 8).

A series of transmissive fractures previously identified at their intersection with the boreholes and characterised through flowmeter tests and optical logging (Le Borgne et al., 2007) can be correlated to reflections in the GPR sections (blue letters in Figs. 8 and 9) based on their dips and extrapolated intersection points given positioning errors of up to 4%. Fractures that could not be correlated to GPR reflections either lie in the mica schist (B1-1) where signal-to-noise ratios are low, have a nearly horizontal dip and hence cannot lead to direct reflections (B1-4 in Fig. 8a: probable phase shift on reflections due to B1-4) or have a too small spatial extent to be seen at $r > 2$ m (B3-2 and B2-2 are only seen up to $r = 4$ m in Fig 8 but are not clearly seen in the 100 MHz data in Fig. 9). A number of hitherto unknown prominent subvertical features can be seen that do not cross the boreholes. As an example there is a larger fracture zone (A3) dipping ~70° that is crossing the sections B1 and B2 at $r$ = 2-10 m.

### 5.2 Cross-hole GPR images

The final stacked and migrated cross-hole sections (Fig. 10) image several features relative to the corresponding CMP locations (for position see Figure 1c for B1-B2 plane). The signal attenuation is much lower at $z > $ ~38 m at which the surrounding rock matrix consist of high-resistive granite. Most features are linear dipping 30-90°, some curved features are shown at $x < $ ~8 m that can be attributed to the elliptic symmetry. Most prominent reflections are subvertical (A3-A7 in Fig. 10).

It is difficult to confidently correlate reflections in the cross-hole image to those identified in the borehole logs since the reflections are imaged as originating away from the cross-hole midpoints. It appears still that certain prominent subvertical reflections seen in the single-hole data can be related to reflections in the cross-hole sections (for example A3 in Figs. 8 - 10).

### 6 Discussion

Multiple-offset data acquisition together with a tailored pre-processing and pre-stack migration made it possible to correlate most transmissive fractures observed in flowmeter and optical televiewer data (Le Borgne et al., 2007) with reflectors observed in the single-hole reflection data (Figs. 8 and 9). In fact, a total of 10 out of 11 transmissive fractures in the granite could be correlated with reflectors. These reflections appear as (sub-)linear features in the final processed and migrated images. Reflectors crossing the boreholes were best identified by first picking reflectors from the migrated 250 MHz data (Fig. 9) followed by verification that these reflectors are visible on individual unmigrated time-sections. The dip angles of the reflectors and those observed in the boreholes were allowed to have a mismatch of up to 10º as fractures/reflectors that appear linear on the scale of the fracture might locally (i.e., where they cross the boreholes) show larger deviations in the dip angle as evidenced at outcrops of similar patterns on the coastline 5 km away. Most transmissive fractures can be correlated with reflectors, but there are also certain reflectors that appear to cross the boreholes that are unrelated to the previously identified permeable fractures. One important result is that we can associate a length scale to previously identified transmissive fractures.

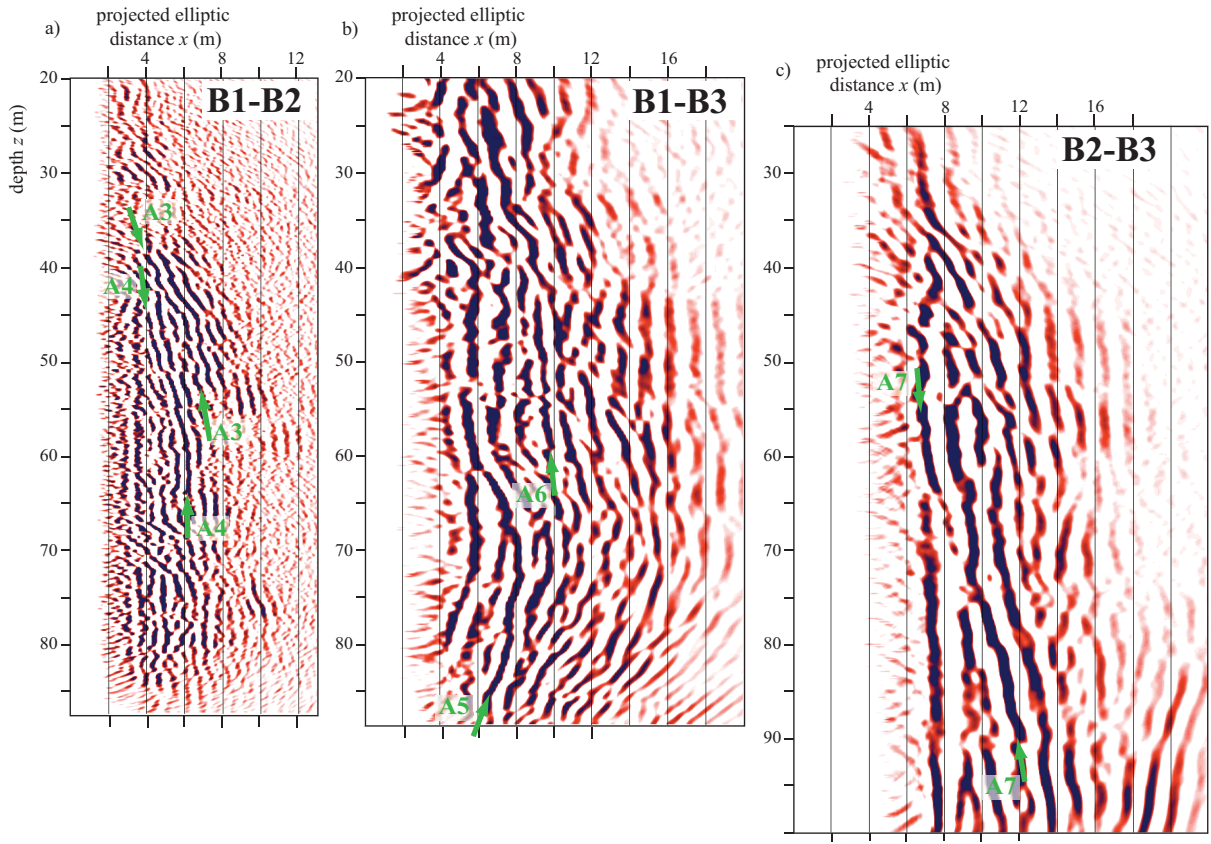

*Fig. 10: Processed and migrated cross-hole GPR sections of (a) B1-B2 (250 MHz), (b) B1-B3 (100 MHz), and (c) B2-B3 (100 MHz). A3-A7 refer to features discussed in the text. The projected elliptic distances x are relative to the CMP locations as shown for B1-B2 in Figure 1c. The axis aspect ratio x:z is 2:1. Note the lack of information at x <~2 m due to the direct wave removal. Note also that dipping individual fractures that are expected to be linear show up as curved features in accordance with Figs. 5b and 5c.*

The migrated single-hole (Figs. 8 and 9) and cross-hole (Fig. 10) data image several prominent fractures or fracture zones with lengths exceeding 40 m. Only one of those (B3-1 in Figs. 8c and 9c) appears to cross the borehole. It is important to note that the anisotropic radiation/reception patterns of the antennas make subvertical dipping features in the single- and cross-hole data more prominent compared to shallow ones because the electric dipole source has its orientation in the vertical direction. Furthermore, imaging limitations at large distances away from the boreholes make it impossible to trace the full length of the more subhorizontal dipping reflectors. It is possible that the previously unknown subvertical fracture zones play an important role in (1) establishing the hydraulic connections observed and (2) providing sustained yield during pumping. This interpretation is supported by Le Borgne et al. (2007) who demonstrate that none of the identified permeable fractures appear to cross more than one borehole. Furthermore, steady-state conditions are established almost instantaneously at this site, which indicates a well-connected fracture network extending over a large scale. Some of these major reflectors can be imaged from different boreholes and with the cross-hole data (Figs. 8-10). Consider A3, as it is imaged in both B1 and B2 single-hole data, and in the B1-B2 cross-hole data, it is possible to better constrain its location (e.g., Olsson et al., 1992; Spillman et al., 2007).

Saline tracer tests monitored with single-hole GPR data will in the future be analyzed to test the conceptual model invoked above. It is also expected that such experiments will decrease the inherent uncertainty caused by projecting 3-D data into 2-D as the injection point is known.

**7 Conclusions**

GPR is one of few geophysical methods that are capable of imaging individual millimeter wide fractures away from boreholes. We have processed and interpreted multiple-offset single-hole and cross-hole GPR data acquired in a granitic rock aquifer. The multiple-offset acquisition not only increased the CMP fold, but also made it possible to image fractures with limited extents and dip angles for which reflections are only visible within a certain offset-range. Our processing scheme allowed us to separate useful reflections from unwanted signals to obtain high-resolution images. Key processing steps included time-zero and geometrical corrections, eigenvector filtering to remove direct wave and ringing effects, careful top muting and migration, and accurate radar wavespeed estimation and borehole positioning by a combined analysis of single- and cross-hole data. The migration (using pre-stack depth migration or DMO-corrections followed by depth migration) was able to handle subhorizontal dips and provided images that were free of major artefacts. The superposition of reflections necessitate a combined interpretation considering both unmigrated and migrated data.

The final GPR sections image a number of reflectors with dips in the range of 30-90° at radial distances of 2-20 m and spatial extents of more than the first Fresnel-zone (2 m at $r$ =20 m down to 0.6 m at $r$ = 2 m). Reflections from a certain reflector can be observed if a normal-vector to the reflector exists which crosses the borehole. We observe the highest resolution in the vicinity of the borehole when using the 250 MHz data, whereas the 100 MHz data are very useful in imaging major subvertical features away from the borehole.

These mostly linear reflections are interpreted as mainly corresponding to fractures, but are also due to other boreholes and the contact zone between mica schist and granite. Ten out of eleven known transmissive fractures in the granite can be correlated to reflections. By identifying the same reflectors in different borehole GPR sections, we can reduce the inherent 360° circular uncertainty of single-hole data and elliptical uncertainty of cross-hole data to some extent. The dip direction of identified fractures remains underdetermined because the three boreholes do not form a triangular prism at depth, but rather lie on a common plane. Prominent sub-vertical reflectors image previously unknown fractures (they do not intersect the boreholes) with lengths exceeding 40 m that may play a key role in determining the flow geometry at the site.

**Acknowledgements**

We are thankful to Alan Green at ETH-Zurich for making the GPR equipment and a processing code available that is supported by Landmark Graphics Corporation via the Landmark University Grant Program. Philippe Pezard at Géosciences Montpellier kindly made the geophysical logging data available. Ludovic Baron at the University of Lausanne designed the radar antenna centralizers. Constructive comments from the Editor Hansruedi Maurer, Stewart Greenhalgh, and two anonymous reviewers helped to improve the manuscript. This research was supported by the Swiss National Science Foundation under grant 200021-124571 and the French National Observatory H+.